\documentclass[twocolumn,showpacs,preprintnumbers,amsmath,amssymb]{revtex4}

\usepackage{graphicx}
\usepackage{epstopdf}
\usepackage{dcolumn}
\usepackage{bm}
\newcommand{\comment}[1]{}
 \begin{document}

\title{Maximizing Influence Propagation in  Networks with Community Structure}

\author{Aram Galstyan$^{1}$}   \email{galstyan@isi.edu} 
\author{Vahe Musoyan$^{2}$} 
\author{Paul Cohen$^{3}$}

\address{ $^1$Information Sciences Institute, University of Southern California, Marina del Rey, CA\\ 
$^2$ Department of Applied Mathematics and Informatics, Yerevan State University, Yerevan, Armenia\\
$^3$ Department of Computer Science, University of Arizona, Tucson, AZ
 }

%


\date{\today}

\vspace{8mm}
\begin{abstract}
We consider the algorithmic problem of selecting a set of target nodes that cause the biggest activation cascade in a network.  In case when the activation process obeys the  {\em diminishing returns} property, a simple hill--climbing selection mechanism has been shown to achieve  a provably good performance. Here we study models of influence propagation that exhibit critical behavior, and where the  property of diminishing returns does not hold.  We demonstrate that in such systems, the structural properties of networks  can play a significant role.  We focus on networks with two loosely coupled communities, and show that the double--critical behavior of activation spreading in such systems has significant implications for the targeting strategies.  In particular, we show that simple strategies that work well for homogeneous networks can be overly sub--optimal, and suggest simple modification for improving the performance, by taking into account the community structure.
\end{abstract}
\maketitle

\section{Introduction}
\label{sec:intro}

Much recent research has focused on understanding how structural properties of networks affect their dynamical behavior \cite{albert2002,dorogovtsev2002,MEJNewman2003,dorogovtsev2008}. For instance, it has been established that critical behavior of epidemic models on random Erd\"{o}s--R\'{e}nyi graphs are absent in certain scale--free networks~\cite{vespignani2001a,vespignani2001b}. This fact has significant implications for a number of important applications. In particular, while scale--free networks are generally robust to random breakdowns, they can be highly vulnerable to intentional attacks that target highly connected nodes~\cite{cohen2001}. This suggests that immunization strategies for such networks should take into account the inherent heterogeneity in the degree distribution. Indeed, it has been shown that  targeted immunization  based on the nodes connectivity hierarchy can significantly lower the networks vulnerability to epidemic attacks~\cite{Vespignani2002}.  

 Here we consider a related problem of {\em maximizing} influence propagation in networks, by targeting certain influential nodes  that have the potential to influence many others. This problem has attracted some recent attention due to potential applications in {\em viral} marketing, which is based on the idea of leveraging existing social structures for word--of--mouth advertising of products~\cite{Domingos2001,Richardson2002,Leskovec2006}. From the algorithmic standpoint, this selection problem can be stated as follows~\cite{Domingos2001,Kempe2003}:  Given a social network, an influence model, and a set of nodes $S$, let $\sigma(S)$ be the expected number of nodes that will be activated by the end of the influence propagation process. Then, for a given budget $n$, the influence maximization problem is concerned with finding the  set $S$ of size $n$ that maximizes the return $\sigma(S)$. While this problem is known to be NP hard for the many influence models, several approximate methods have been developed. An important result established in ~\cite{Kempe2003} states that for a class of models for which the return function is sub--modular,  a simple hill--climbing algorithm, which works by greedily selecting the next best candidate node,  yields a solution which is  guaranteed to be within $\sim 63\%$ of the optimal. Sub--modularity of the return function means  that   $\sigma(S \cup \{\omega\})-\sigma(S) \ge \sigma(T \cup \{\omega\})-\sigma(T) $ for any node $\omega$ and any $S \subseteq T$. In other words, the expected return for targeting a node diminishes with the number  of targeted nodes.

While it is quite safe to assume that the   diminishing returns property is satisfied in saturated, or near--saturated,  markets, those models might fail to capture the dynamics of emerging markets, where the condition of the sub--modular growth can be violated. Indeed, many economical and social phenomenon are better described in terms of critical phase transitions, where a huge growth is observed only after some threshold conditions are met.  Here we are interested in this latter case. Our main result is that in such critical systems, the structural properties of networks can play a significant role  in the dynamics of the influence propagation. Consequently,  selection strategies that do not account for those structural properties  might produce vastly sub--optimal results.

To be more specific, let us focus on the so called linear threshold models (LTM)~\cite{Granovetter1978,watts2002} where a node is activated whenever the fraction of its active neighbors exceeds some pre-defined threshold,  $\sum_{j\in {\mathcal N}_i} w_{ij} \ge \theta_i$. Here ${\mathcal N}_i$ is the set of active neighbors of node $i$, $w_{ij}$ is the normalized weight of the link between the nodes $i$ and $j$, $\sum_jw_{ij}=1$, and $0<\theta_i<1$ is the activation threshold for the node $i$. Usually, $\theta_i$-s are assumed to be random variables  reflecting the uncertainty about individuals.  \comment{such a choice of the threshold distribution might actually prohibit the criticality in the activation dynamics for dense networks. To understand why this is the case,  let us call a node {\em vulnerable} if it needs only one active neighbor to be activated itself~\cite{watts2002}. Then, note that for sufficiently dense networks, there will be a percolating cluster of nodes that have a threshold below that need only one active neighbor  be activate The reason for this is the emergence of a percolating cluster of vulnerable nodes Below we will consider modified models that allow for critical behavior.}

\begin{figure}[!t]
 \center
\begin{tabular}{c}
(a)\\
\includegraphics[width=0.35\textwidth]{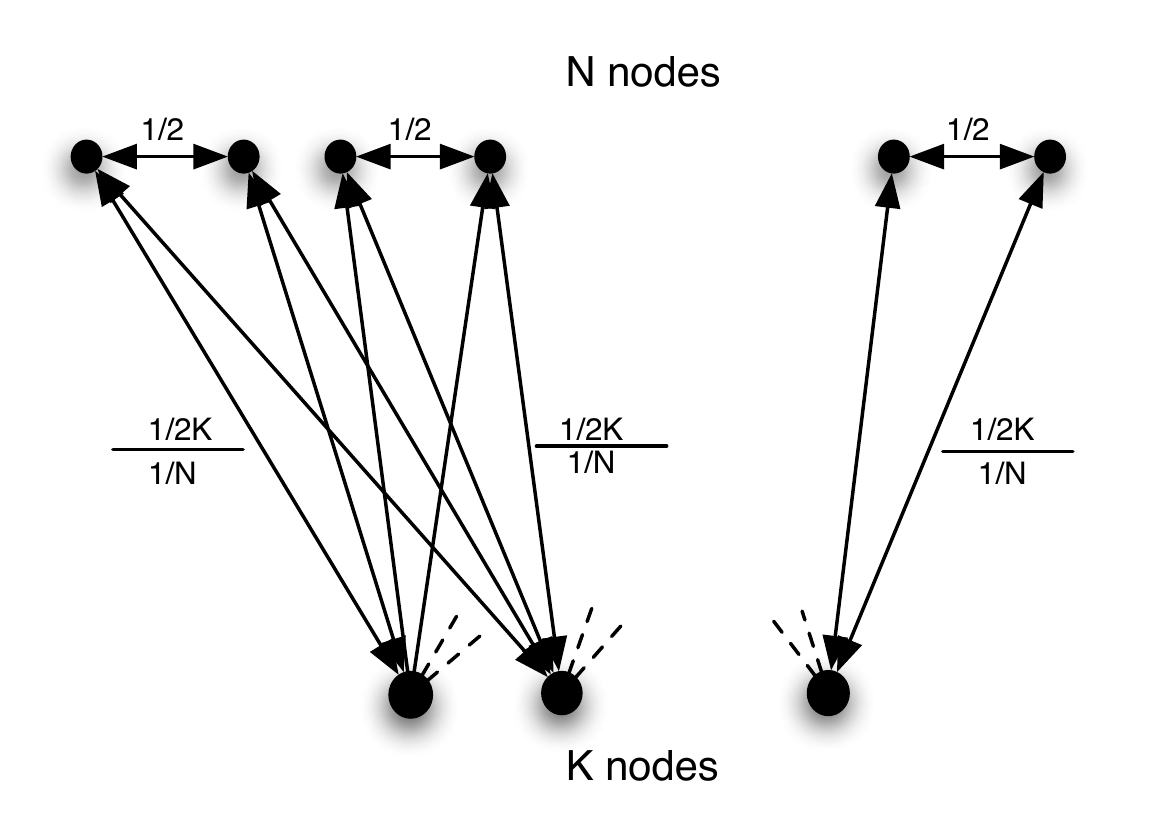}\\
(b)\\
\includegraphics[width=0.35\textwidth]{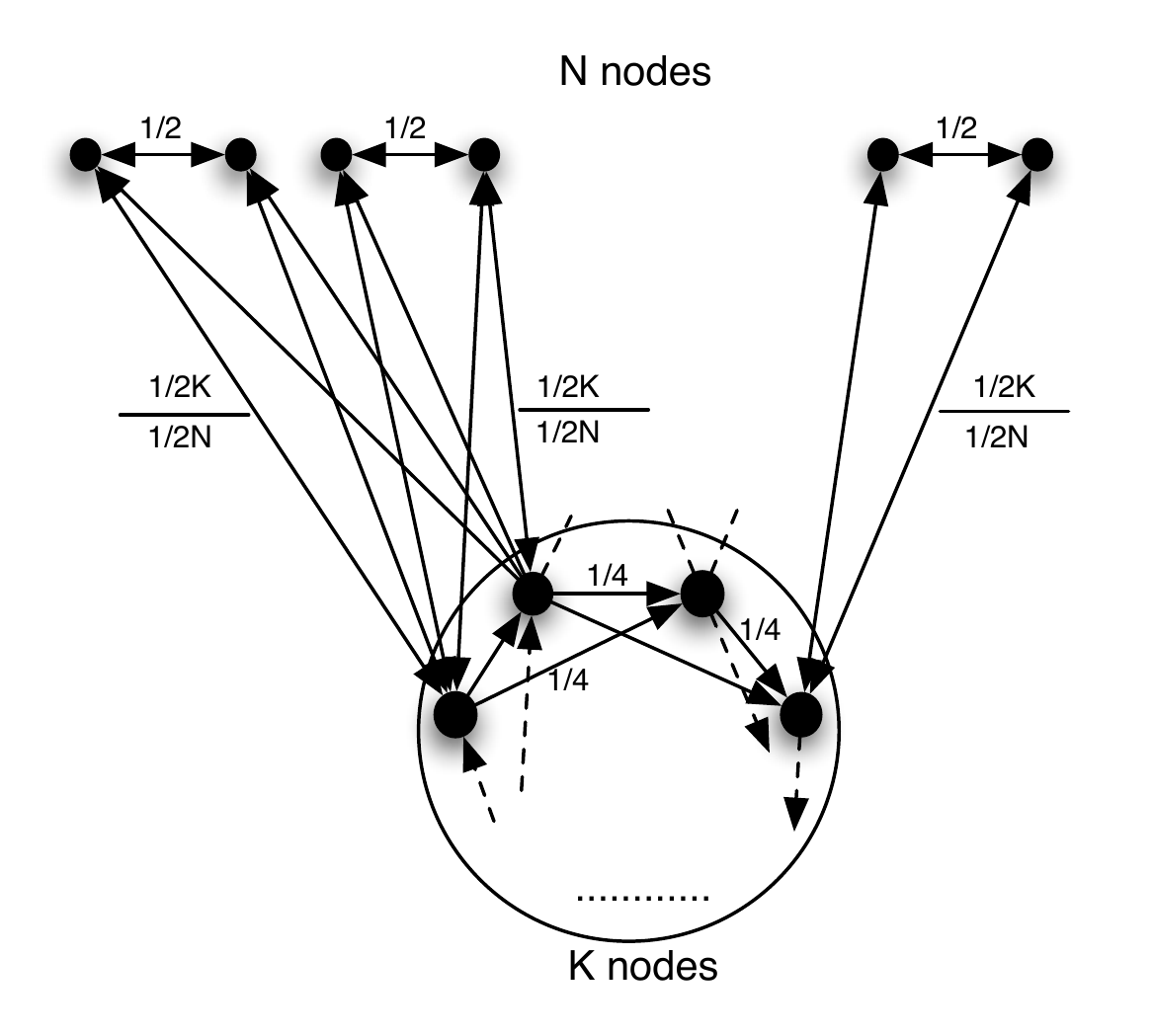}
\end{tabular}
\caption{ Two hypothetical networks illustrating the limitation of the hill--climbing algorithm. In both networks, each node from the bottom (upper) row affects all the upper (lower) nodes with weight $1/2K$ ($1/N$). In (b), each of the $K$ nodes in the lower row is  influenced by its two preceding  neighbors with weights $1/4$. }
\label{fig:1} 
\end{figure}

Consider  the unfolding of the LTM dynamics on the hypothetical influence graphs depicted in Figure~\ref{fig:1}. In Figure~\ref{fig:1} (a), the nodes in the upper row are connected in pairs, and each link has a weight $1/2$. Each node from the bottom (upper) row affects all the upper (lower) nodes with weight $1/2K$ ($1/N$). Assume  fixed thresholds $1/2$ for all the nodes .  If one follows the hill--climbing algorithm, then it is easy to see that only the upper nodes  will be selected.  Thus, after targeting $n$ nodes, the number of active nodes is exactly $2n$. If $n<K$, then this is indeed an  optimal solution to influence maximization problem. However, for $n\ge K$this solution is clearly sub--optimal as targeting the bottom $K$ nodes would activate all the $N+K$ nodes in the network.  The sub--optimality is even more dramatic for the network depicted in Figure~\ref{fig:1} (b), which is obtained from Figure~\ref{fig:1} (a)  by adding $2K$ links so that each of the lower nodes is now influenced by its  two proceeding neighbors with weights $1/4$. Assuming a threshold $1/2$ for all the nodes,  one observes that the greedy selection policy will again result in a final active set of size $2n$. However, a simple inspection shows that if one activates two neighboring nodes from the bottom row, then it will cause a global cascade among the lower nodes, which will consequently propagate to the upper nodes and activate them as well. This suggests that for large $N$, greedy selection mechanism produces  vastly sub--optimal solution. 

\begin{figure}[!h]
 \center
\includegraphics[width=0.34\textwidth]{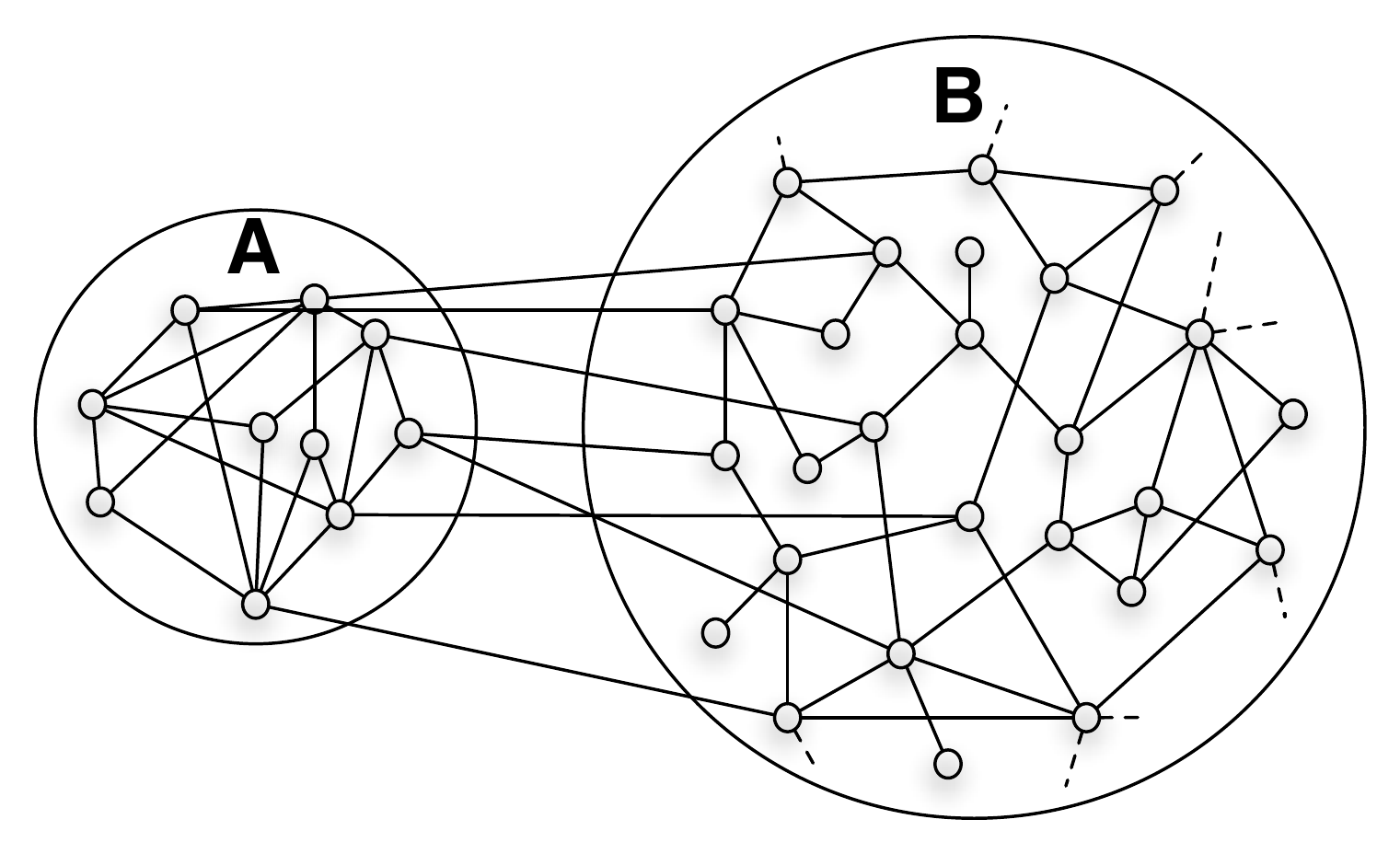} 
\caption{Schematic illustration of a bi--community network.}
\label{fig:graph}
\end{figure}

While  the examples above seem peculiar, the main claim of our paper  is that the underlying effect is rather  general and present in more realistic  models as well. Indeed, the two contributing factors to the behavior described above are the  {\em critical nature of the activation dynamics}, and the {\em structural heterogeneity of the network}. The criticality is manifested by the fact that there is a threshold number $n_c$ so that for $n<n_c$ influence propagation is localized, whereas for   $n\ge n_c$ the activation spreads throughout  all (or almost all) the nodes in the network.  And by the structural heterogeneity we mean different and heterogeneous linkage--patterns among the nodes. A large class of networks that fit this description are networks with well--defined  {\em communities}~\cite{girvan2002,M.E.J.Newman06062006}.  In particular, here we focus on networks that are composed of a relatively small, tight community that is connected with a  larger population of nodes (see Figure~\ref{fig:graph}). 

\vspace{-5mm}

\section{Activation dynamics on bi--community Erd\"{o}s--R\'{e}nyi graphs}

We have previously analyzed activation dynamics on networks  composed of two loosely coupled Erd\"{o}s--R\'{e}nyi graphs~\cite{galstyan2007_PRE}.  To make this paper self--contained, we below provide our analysis. Instead of using the traditional linear threshold model, we focus on a modified version, where the threshold condition is applied not to the fraction of active neighbors, but their number. Thus, a node is activated whenever the number of its active neighbors is greater or  equal a predefined threshold, $h$. The reason for modifying the model is that the consequent analysis is simple. Furthermore, one can argue that the modified model might more plausible from the social choice standpoint: Indeed, it is hard to imagine that, while making a decision based on the opinion of few friends, one ``weighs" the advice by the overall number of friends.  We stress, however, that our main results are valid for the fractional threshold model as well, provided that it possesses critical behavior.

 Let us first focus on a single Erd\"{o}s--R\'{e}nyi graph with an average connectivity $z$. Let $P_h$ denote the fraction of nodes with threshold $h$, and let $\rho_{0}$ be the budget, {\em i.e.}, the fraction of targeted (initially activated) nodes. In this section we consider random uniform targeting, so that each node has a probability $\rho_0$ to be targeted.  Following the same line as in~\cite{galstyan2007_PRE} it can be shown  that the fraction of activated nodes at the end of the cascading process satisfies the following transcendental equation:  
\begin{equation}
\rho^s= 1  -  (1 - \rho_0)\sum_{h=0}^{\infty}P_h Q(h; z\rho^s)
\label{eq:steadystate}
\end{equation}
where $Q(n,x)=\sum_{k<n}e^{-x}x^k/k!$ is the regularized gamma function. To understand this expression, note that in the steady state, a node with a threshold $h$ is inactive if it is connected with less than $h$ active neighbors, and it is not among the initially targeted set. The former happens with probability   $Q(h; z\rho^s)$, while for the latter this probability is  $(1 - \rho_0)$, hence yielding Equation~\ref{eq:steadystate}.

For a fixed connectivity $z$, the solution of the Equation~\ref{eq:steadystate} depends on the budget $\rho_0$, as well as on the threshold distribution function $P_h$. Let us elaborate on the latter dependence in more details. First of all, we assume that $P_0 = 0$, {\em i.e.}, there are no nodes that activate spontaneously, aside from the initially targeted nodes. Furthermore, simple inspection shows that the dynamical properties of the model depend on the fraction of nodes with threshold $h=1$, $P_1$. Following~\cite{watts2002}, we call these nodes {\em vulnerable} since they will activate whenever one of their neighbors is active. Clearly, if the fraction of the vulnerable nodes is sufficiently large, a single node might trigger a global cascade throughout the network. In particular,  a global cascade will happen whenever the  vulnerable nodes form a giant connected component~\cite{watts2002}, which, for the random Erd\"{o}s--R\'{e}nyi graphs translates into $P_1z=c> 1$. Below we consider the case when $P_1$ is either zero, or sufficiently small, $P_1\ll 1/z$, so that for a network of size $N$, the number of nodes required to cause a global cascade must be of order $O(N)$ as $N \rightarrow \infty$.

For the latter case, the analysis of Equation~\ref{eq:steadystate} yields the following observation: For a given connectivity $z$, there is a critical fraction $\rho_c$ such that for $\rho_0 < \rho_c$ the activation process is localized, while for $\rho_0 >\rho_c$ activation spreads to all the nodes in the network. One can obtain the following expression for the critical density:
\begin{equation}
\rho_c = 1-  \biggl [  ze^{-x_0}\sum_{h=0}^\infty P_{h+1} \frac{x_0^{h-1}}{(h-1)!} \biggr ] ^{-1}
\label{eq:rho_c}
\end{equation}
 where $x_0$ satisfies the following equation:
\begin{equation}
1 - \frac{x_0}{z} = \frac{\sum_{h=0}^\infty P_{h+1} \frac{x_0^{h-1}}{(h-1)!}}{\sum_{h=0}^\infty (1 - D_{h}) \frac{x_0^{h-1}}{(h-1)!}}
\label{eq:x0}
\end{equation}
Here $D_h = \sum_{i\le h} P_i$ is the cumulative distribution function for the activation thresholds. 


Consider now the activity spreading in two coupled Erd\"{o}s--R\'{e}nyi networks of sizes $N_a$ and $N_b$ as depicted in Figure~\ref{fig:graph}, with connectivities $z_{aa},z_{bb}$ within the groups, and $z_{ab}=z_{ba}N_b/N_a$ across the groups. Assume that the cascading process in group $A$ is not affected by  cross-group
links, so that the activation for $A$ nodes is governed by the Equation~\ref{eq:steadystate}. For the $B$ nodes, the activation dynamics is given by a similar equation, with the only difference that it is affected by the presence of active $A$ nodes, and the steady state fraction of active $B$ nodes  satisfies the following equation:
\begin{equation}
\label{eq:steadyb}
\rho_b^s = 1  -  (1-\rho_{b,0}) \sum_{h=0}^{\infty}P_h Q(h; z_{bb}\rho_b^s + z_{ba}\rho_a^s)
\end{equation}

where  $\rho_a^s$ is the steady state fraction of active $A$ nodes. Thus, the presence of active $A$ nodes facilitates the activation of $B$ nodes, and the effect depends on the across the group connectivity $z_{ba}$.  Specifically, if $z_{ba}$ is very small, then, in order to achieve a global activation in group $B$, one needs to target fraction of $B$ nodes above a certain threshold $\rho_{b,c}$. However, even below the threshold, there is a possibility of a global cascade in group $B$ if the across the group connectivity $z_{ba}$ is sufficiently large. Indeed,  a simple analysis shows\cite{galstyan2007_PRE} that for a fixed within--group connectivity $z_{bb}$,  there is  a critical connectivity $z_{ba}^c$ so that for $z_{ba}>z_{ba}^c$ the activation will propagate from group $A$ to group $B$ and cause a global cascade. 

\subsection{Influence Maximization in bi--community Erd\"{o}s--R\'{e}nyi graphs}

The analysis above suggest that discarding the community structure might result in sub--optimal solution to the influence maximization problem. Indeed, since the critical number of nodes necessary to cause a cascade for a given connectivity grows linearly with the network size, it might be more beneficial to target the smaller group first and cause an activation cascade in that group. Afterwards, the activation will propagate through the larger network, provided that the density of  links between the groups is sufficiently strong.  Strictly speaking, the analysis above applies to the random targeting strategies. However, one might expect a similar reasoning to hold for the greedy selection heuristics as well.  Indeed, below we validate this hypothesis for synthetic random graphs, which which are similar to those used in the evaluation of community finding algorithms~\cite{Danon2005}. Namely, we assume that the network is composed of $L$ groups, with $N_L$ nodes in each. Each pair of nodes within and across the same groups are linked with probability $p_{in}$ and $p_{out}$, respectively, with corresponding connectivities $z_{in}=p_{in}N_m$ and $z_{out}=p_{out}(N-N_L)$. We assume that one of those  $L$ groups constitute the group $A$, while the remaining $L-1$ communities form  $B$.

We tested the greedy selection algorithm for  with both integer and fractional versions of the linear threshold model. The conventional greedy selection works as follows: Starting from the empty set $S =\emptyset$, the algorithm finds a single node that causes the largest cascade, adds it to $S$ (if there are many candidates nodes, then one needs a tie--breaking mechnaism: Here we choose the node with the highest connectivity). This process is then repeated  $n$ times. We compared this simple hill--climbing scheme with another one,  which works exactly the same way, but now the candidate nodes are selected {\em only} from the smaller community $A$.  In the following, we differentiate the latter algorithm by a subscript $A$. 
\begin{figure}[!t]
\begin{tabular}{c}
\includegraphics[width=0.4\textwidth]{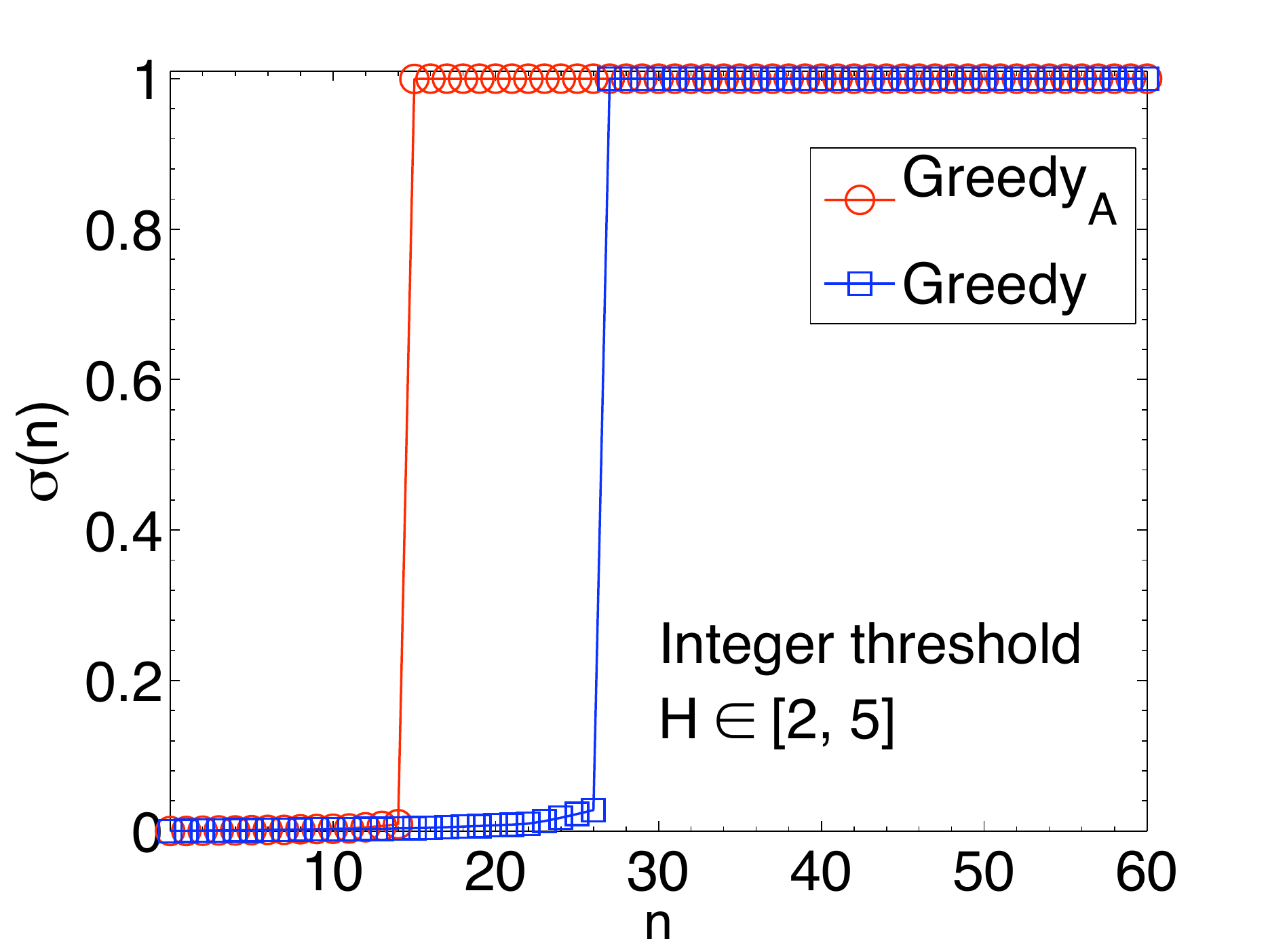}\\
\includegraphics[width=0.4\textwidth]{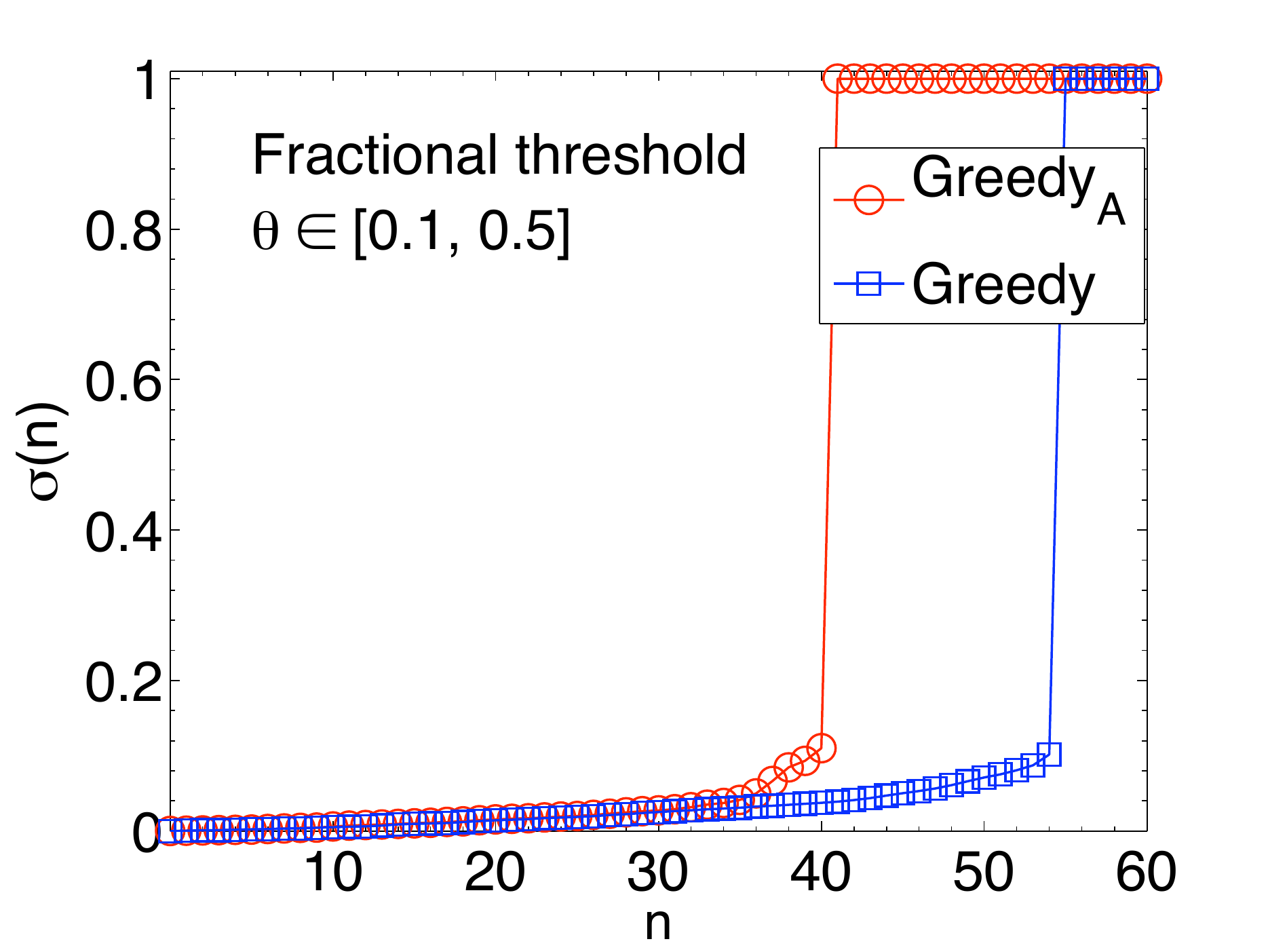}
\end{tabular}
\caption{Comparison of target--selection strategies for integer (upper panel) and fractional (lower panel) threshold models. We used $L=10$, and $N_L=500$, and the total network size is $N=5000$. The connectivities were set to $z_{in}=z_{out}=8$, and the thresholds were chosen randomly (uniformly)  from the intervals shown in the inset. }
\label{fig:3} 
\end{figure}

In Figure~\ref{fig:3} (a) we plot the fraction of activated nodes against the budget $n$, for  the two selection strategies, and for the integer threshold model. The  connectivities are set to $z_{in}=z_{out}=8$, and the thresholds were chosen randomly and uniformly from the interval $[2,5]$. One can see that the strategy of targeting nodes from the smaller community  is generally more efficient, as it achieves a global cascade with a significantly lower budget $n$. More precisely, for small and large values of $n$,  both methods have a similar performance. However, there is a window $[n_{A}^c, n_{B}^c]$, within which the selection of $A$ nodes is clearly superior. Recalling the analysis from the previous section, it is clear that $n_A^c$ corresponds to the critical threshold for which the activation spreads throughout  group $A$, and then spills into the rest of the network. If one targets nodes from the general population, on the other hand, this critical effect does not come into play until later, when larger number of nodes have been selected.  

The same picture holds for the fractional--threshold mode as shown in Figure~\ref{fig:4} (b), where we again compare both algorithms. The fractional thresholds are uniformly distributed on the interval $[\theta_{min}, \theta_{max}]$, with $\theta_{min}=0.1$, $\theta_{max}=0.5$ used here. Again, we observe that for both small and large values of $n$, both methods have a similar performance, while for an intermediate values of $n$, the strategy that selects $A$ nodes are superior.     

\section{Influence Maximization  on Scale--free graphs}

Real--world networks have statistical characteristics that significantly deviate from the Erd\"{o}s--R\'{e}nyi model. In particular, many networks exhibit power--law degree distribution. Below we examine influence maximization in such networks. We show that the sub--optimality of the simple greedy algorithm persists whenever the connectivity variance is limited, so that activation dynamics still demonstrates critical behavior, 

Let us first focus on the activation dynamics in a single population. Furthermore, for the sake of simplicity, let us assume that all the nodes have the same activation threshold $H$. We consider a network with degree distribution $p_k=ck^{-\gamma} $, $m\le k\le M$, where $m$ and $M$ are lower and upper cutoffs, respectively, and $c$ is a normalizing constant,
\begin{equation}
c = (\gamma -1)\frac{m}{1-\beta^{\gamma}} ,\ \beta = m/M
\end{equation}
Let    $\rho_k$ be the fraction of nodes with connectivity $k$  that are active at the end of the activation process, and let $\rho_{0,k}$ be the fraction of targeted $k$--nodes. Using similar arguments as in the Erd\"{o}s--R\'{e}nyi  case, one can show that $\rho_k$--s satisfy the following equations ($k=m,m+1,..M$):
\begin{equation}
\rho_k = 1  - (1 - \rho_{0,k})\sum_{m=0}^{H-1}P(m|k).
\label{eq:dynsf}
\end{equation}
Here $P(m|k)$ is the probability that $m$ out of $k$ edges leaving from a vertex point to an activated node. Let $\theta$ be the probability that a randomly chosen edges leads to an active node.  Then $P(m|k)$ is a binomial distribution that for large $k$ can be approximate by the Poisson distribution with a mean $\theta k$. Furthermore, for uncorrelated networks considered here, $\theta$ can be written as  
\begin{equation}
\theta = \frac{\sum_k k p_k \rho_k}{\sum_k k p_k } \equiv \frac{1}{z} \sum_k k p_k \rho_k ,
\label{eq:theta}
\end{equation}
 where $z=\sum_k k p_k$ is the average connectivity. To understand this expression, note that the probability that a randomly chosen edge leads to a node with degree $k$ is proportional to $kp_k$ (for uncorrelated networks), and the probability that this node will be active is simply $\rho_k$. 

Combining Equations~\ref{eq:dynsf} and ~\ref{eq:theta}  we obtain the following self--consistent equation for $\theta$ in the continuos approximation
\begin{equation}
\theta = 1 - \frac{c}{z}\int_{m}^{M} dk (1-\rho_{0,k})\sum_{j=0}^{H-1}e^{-k \theta} \frac{(k\theta)^{j+1-\gamma}}{j!}
\label{eq:steadystate_theta}
\end{equation}

To proceed further, we need to specify the targeting function $\rho_{0,k}$. We considered two cases -- random selection  $\rho_{0,k}=\rho_0 = const$; and maximum degree (MD) selection heuristics $\rho_{0,k}=\Theta(k-m_0)$, where $\Theta$ is the step function, and the cutoff $m_0$ is found from the budgeting constraint $\int_{m_0}^M p_k=\rho_0$, which yields 
\begin{equation}
m_0=m\biggl [ {\beta^{\gamma-1} + \rho_0(1-\beta^{\gamma-1})} \biggr ]^{-\frac{1}{\gamma-1}}
\label{eq:cutoff}
\end{equation} 

\begin{figure}[!h]
\includegraphics[width=0.45\textwidth]{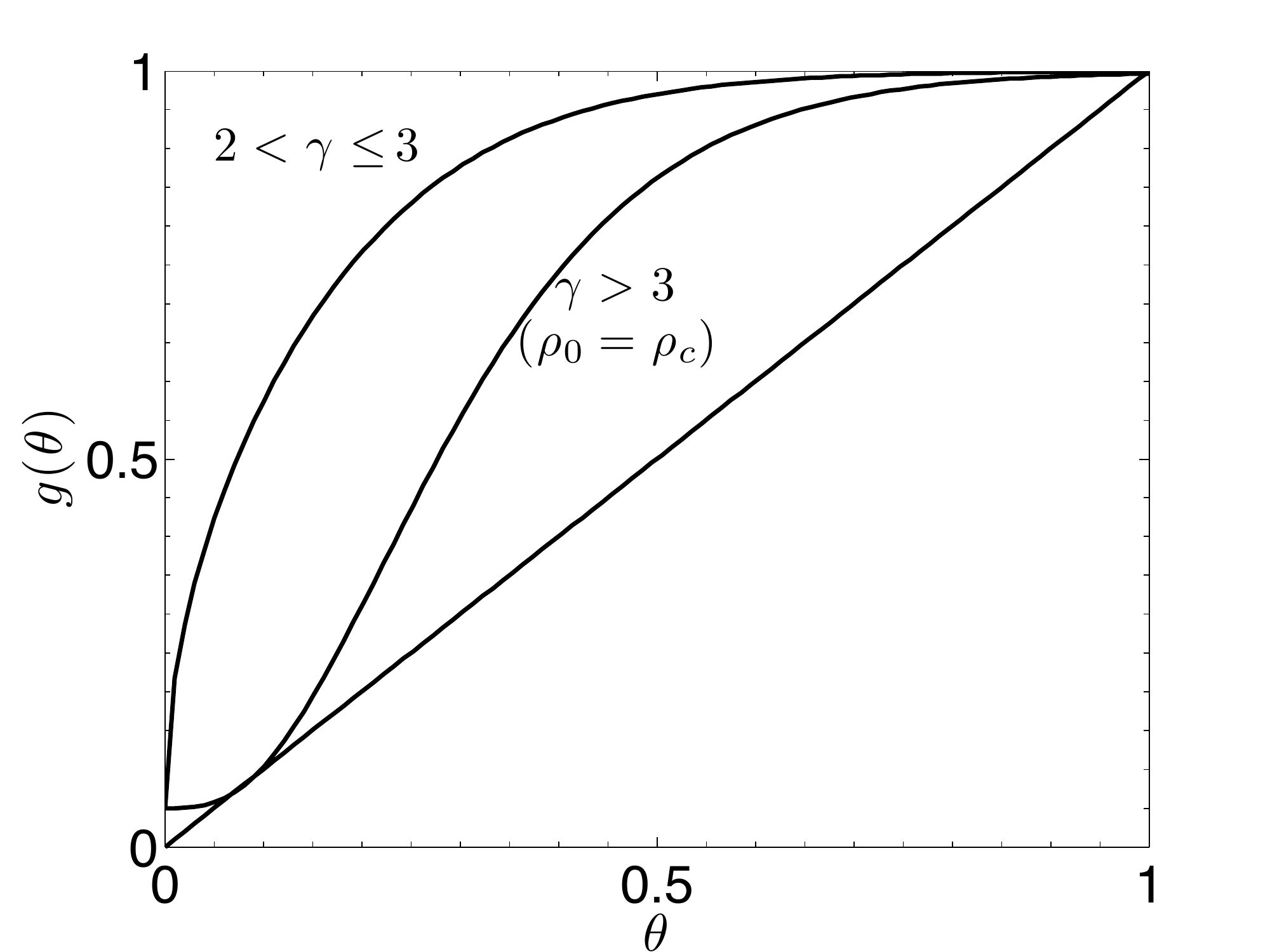}
\caption{Graphical illustration of the steady state equation for the random targeting strategy. }
\label{fig:4} 
\end{figure}

Examination of  Equation~\ref{eq:steadystate_theta}  for both strategies can be summarized as follows. First of all, it is easy to see that for sufficiently dense networks, $\theta = 1$ is always a solution. Thus, for  sufficiently large $\rho_0$ the steady state corresponds to a fully activated network. Furthermore, for $\gamma>3$, there is a critical fraction $\rho_c$ below which another solution appears, as shown schematically in Figure~\ref{fig:4}. In the region $2<\gamma \le 3$, the critical behavior is suppressed if there is no upper connectivity cutoff, $\beta=0$. Namely, for any finite $\rho_0$, the network is fully activated at the end of the cascading process, for arbitrary activation threshold $H$. This is due to the infinite second moment of the connectivity distribution.  The criticality is recovered, however, if one introduces an upper cutoff. More details are provided in the Appendix. 

\begin{figure}[!h]
\begin{tabular}{c}
\includegraphics[width=0.45\textwidth]{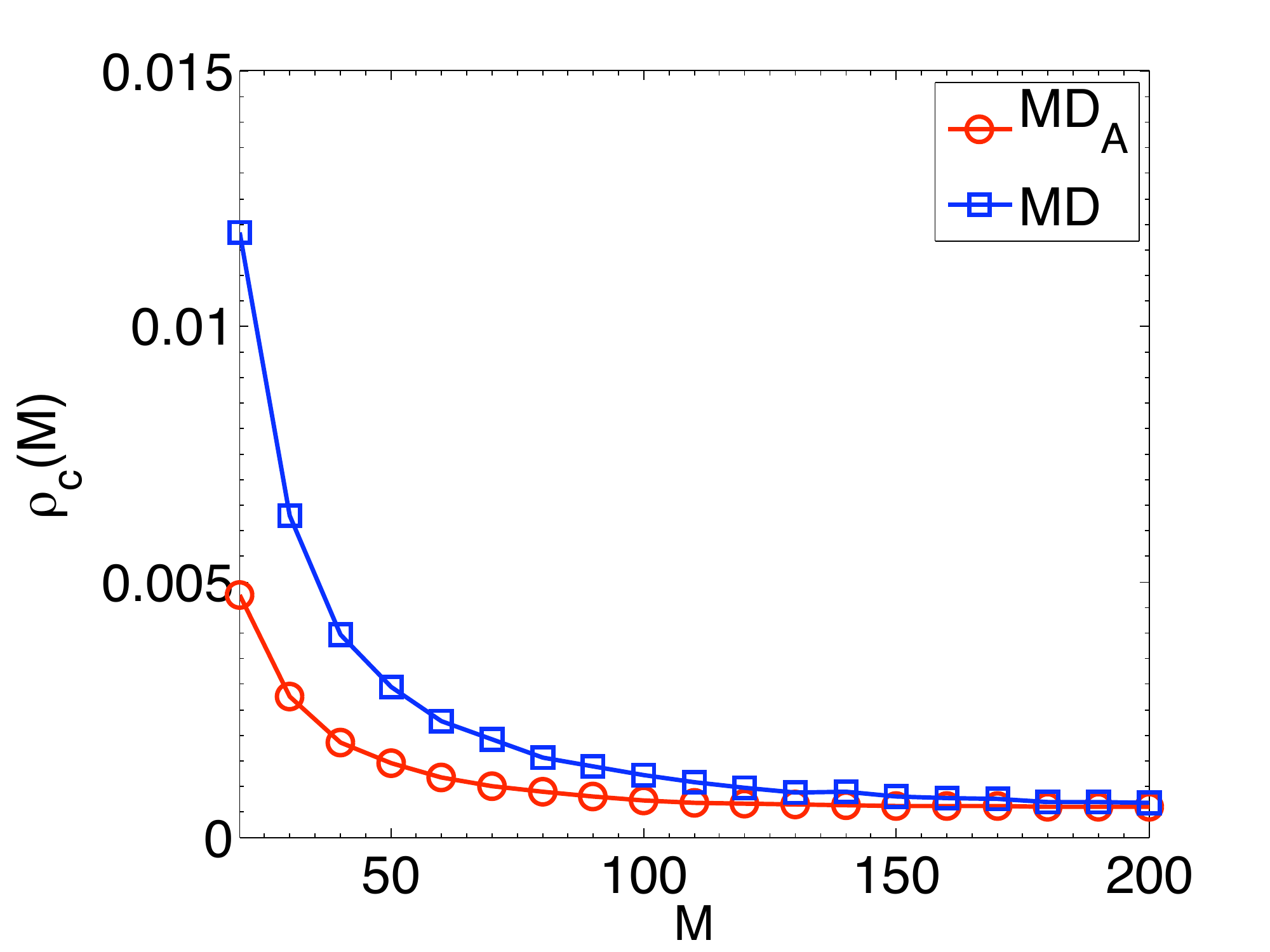}\\
\includegraphics[width=0.45\textwidth]{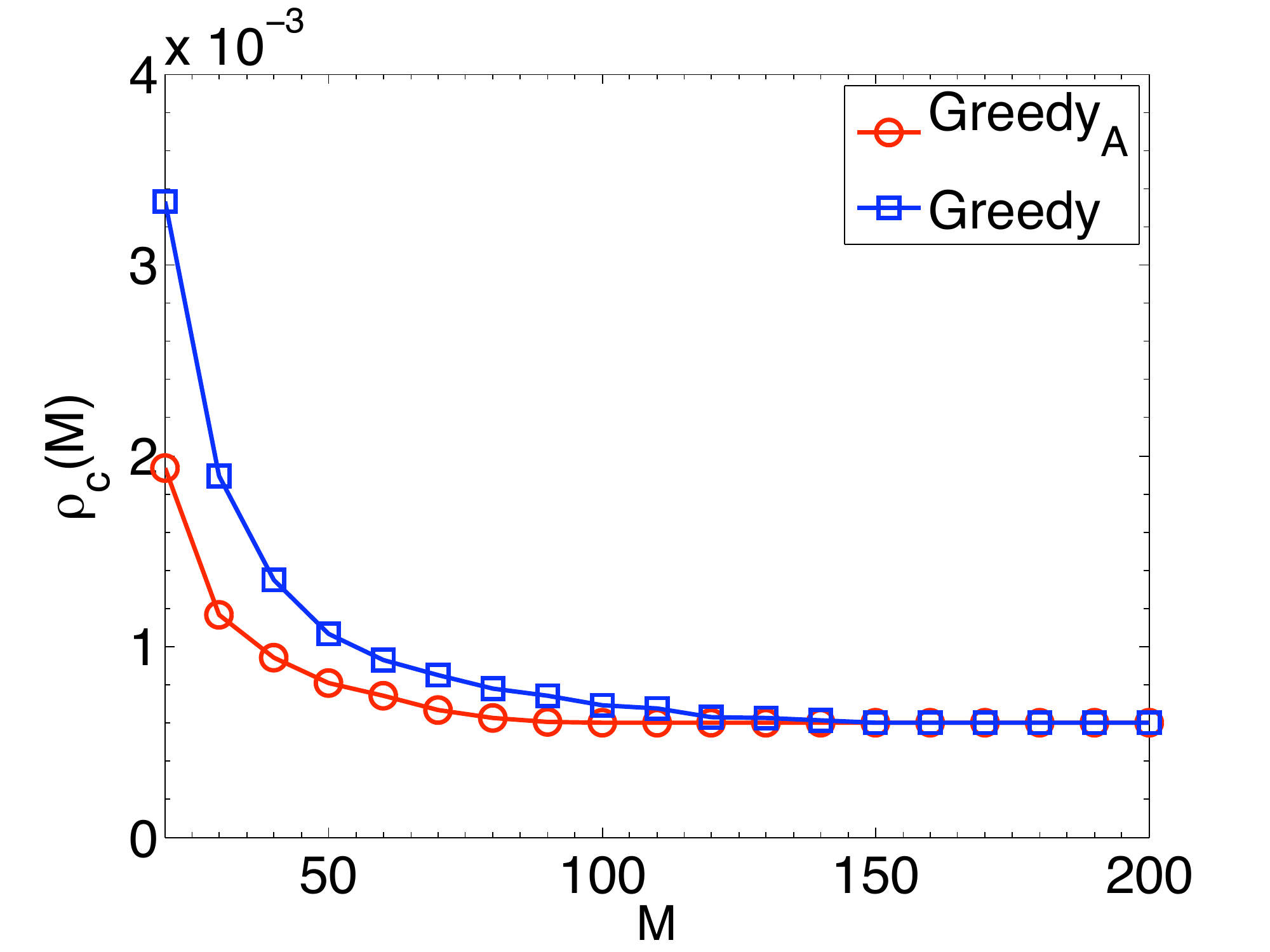}
\end{tabular}
\caption{Comparison of target--selection strategies for scale--free network with $\gamma =2.5$. We used $m=10$, and $N_a=1000$, and the total network size is $N=5000$. }
\label{fig:5} 
\end{figure}

Next, we examine the predictions of the above analysis for the influence maximization problem in networks with power--law degree distribution.  The networks were generated according to the configuration model~\cite{Molloy1995,MEJNewman2001}, with a slight modification to account for a community structure. Namely, we assigned each node to one of two communities. Then, if the generated candidate edge was linking nodes in different communities, that edge was rejected with probability $p_{in}$. Thus, $p_{in}=0$ corresponds with a single--community scale--free network, while $p_{in}=1$ corresponds to two completely disjoint networks. We choose $p_{in}$ such that  the network has a well--defined community structure, but at the same time, the number of links across the communities is sufficiently large so that the cascade can spread from one community to the other.

We examined the impact of community structure on influence maximization for several targeting strategies and for varying connectivity cutoff $M$.  For relatively smaller cutoffs ($M\sim m$), the behavior of the activation spreading should be similar to the results for the random Erd\"{o}s--R\'{e}nyi graphs, while for larger $M$ it recovers the scale free characteristics.  In Figure~\ref{fig:5} (a) and (b) we compare the two variants of the maximum degree heuristics and the greedy strategy, respectively, by plotting the minimum expected budget one needs to achieve a global cascade in the network. The subscript $A$ mean that the corresponding heuristics is applied to the nodes from the community $A$. In the results presented below we used communities of size $N_a=1000$, $N_b=4000$, and $p_{in}=0.8$, and the degree distribution is characterized by $\gamma=2.5$, and the lower cutoff is $m=10$.  Each point was averaged for $100$ random trials. One can see that for small $M$, the difference between two targeting strategies are indeed significant, similar to the results in random Erd\"{o}s--R\'{e}nyi graphs. Note that  in absolute terms, the difference for the maximum degree heuristics is significantly larger compared to the greedy targeting case. This difference however, diminishes as one increases the cutoff. Thus, for networks with very large cutoffs, targeting nodes from the smaller group does not provide any improvement. In fact, depending on the actual cutoff mechanism, one might be better of targeting nodes from the larger community.  For instance, for networks with a power--law exponent $\gamma$, the so called {\em natural} connectivity  cutoff scales with the network size  as $\sim N^{1/(\gamma-1)}$~\cite{dorogovtsev2002}: Thus, one might argue that larger community should have a higher cutoff, which might make it more beneficial to target nodes from that community.

\section{Discussion}
In this paper we examined the problem of maximizing influence propagation in structured heterogeneous networks. We demonstrated that  for models with critical behavior, the structural properties of the network, and specifically, its community structure, can have important implications for the influence maximization problem. We demonstrated analytically that for two--community networks,  targeting nodes from the smaller community might cause a global cascade with significantly fewer number of seed nodes. This effect becomes increasingly important if the sizes of two communities are vastly different.  We also showed through numerical simulations that a similar picture holds for multi--community networks. 

 In practice, one of course does not have precise estimates of model parameters, such as activations thresholds, or even the precise topology of the network. Thus, the problem of finding the optimal target set for influence maximization might not be well defined. On the other hand, with the surge in online networking sites, information about underlaying community structure in such networks is often available. Our results suggest that in such cases, paying attention to the community structure of the network might be beneficial for influence maximization.

We also note that the networks considered here mimic scenarios where  innovations are introduced through a small community of early adopters.  In this respect, our work resonates well with organizational viscosity model of Krackhardt\cite{Krackhardt1997,McGrath2003} that describes diffusion of ideas in an organization. Here  organization is modeled  as a number of interacting sub--units, with closer social ties within each unit. When the organization has a more or less homogeneous structure, then a newly introduced idea  cannot survive unless it is initially adopted by a large number of individuals. However,   if the network describing the interaction of sub--units meets certain structural  conditions, then the idea might take over the whole population even starting from a small number of initial adopters. 

\begin{acknowledgments}
This research was partially supported by the U.S. ARO MURI grant W911NF--06--1--0094.
\end{acknowledgments}

\appendix*
\section{}

We first focus on the random targeting case. Let us define $\delta=\gamma-2$, and $\beta = m/M$, and $x=m\theta$. Then the steady state Equation~\ref{eq:steadystate_theta} corresponds to the zeros of the following function: \begin{equation}
g(x) = 1-\frac{x}{m} - \frac{1-\rho_0}{1-\beta^{\delta} } \delta x^{\delta} \int_{x }^{x/\beta }dte^{-t} \sum_{j=0}^{H-1} \frac{t^{j-\delta-1}}{j!}
\label{eq:dgdx2}
\end{equation}
A simple inspection shows that $x=m$ is always a solution (aside from exponentially small corrections). For $\delta>1$ ($\gamma>3$), there is a critical fraction $\rho_c$ below which other solution appears, as it is schematically shown  in Figure~\ref{fig:4}. Thus, we need to show that for $\delta>1$, Equation~\ref{eq:dgdx2} has a solution for small $x$.  Note that derivative of $g(x)$ is given by 
\begin{equation}
g^{\prime}(x) = -\frac{1}{m}+\frac{1-\rho_0}{1-\beta^{\delta} } \frac{\delta x^{\delta-1}}{(H-1)!} \int_{x }^{x/\beta }dte^{-t}t^{H-1-\delta}
\label{eq:dgdx3}
\end{equation}

Consider the case  $\beta = 0$ ($M \rightarrow \infty$).  Starting at $g(x=0)=\rho_0$, the function $g(x)$ is a strictly decreasing over the interval $0<x<x_0$,  where $x_0$ is determined from  $g^{\prime}(x_0)=0$, which yields 
\begin{equation}
(1-\rho_0) \frac{\delta x_0^{\delta-1}}{(H-1)!} \int_{x_0 }^{\infty }dte^{-t}t^{H-1-\delta}=\frac{1}{m}
\label{eq:dgdx4}
\end{equation}
There are two separate cases: For $H-\delta>0$, the integral, for small $x_0$, can be replaced  by the gamma function $\Gamma(H-\delta)$, which yields 
\begin{equation}
x_0^{\delta-1} \approx \frac{(H-1)!}{m\delta \Gamma(H-\delta)}\frac{1}{1-\rho_0}
\label{eq:x0a}
\end{equation}
And for $H-\delta<0$, after integrating by parts and keeping the leading term we obtain
\begin{equation}
x_0^{H} \approx \frac{(\delta - H)(H-1)!}{m\delta}\frac{1}{1-\rho_0}
\end{equation}
In both cases, $x_0$ remains finite as $\rho_0\rightarrow 0$. Thus, for sufficiently small  $x$, one has $g(x)\approx \rho_0-cx$, where $c>0$ does not depend  on $\rho_0$. Consequently, $g(x)$ will intersect zero for sufficiently small $\rho_0$. More precisely, the critical fraction $\rho_0=\rho_c$ for which the other solution appears is found from $g(x_0)=0$.

 Now consider the case $0<\delta<1$, (or $2<\gamma<3 $). Without connectivity cutoff ($\beta = 0$) the integral in~\ref{eq:dgdx3} $x$ for small $x$ does not depend on $x$, and can be approximated by $\Gamma(H-\delta)$. Thus, the derivative $g^{\prime}(x)$ behaves as  $g^{\prime}(x)\sim 1/x^{1-\delta}$ for small $x$. Consequently, there is no other solution except for the one at $x = m$. A similar  argument holds for $\gamma = 3$, where $g^{\prime}(x)$ remains finite, but positive for small $x$. The situation changes as one introduces a finite cutoff $M$. Indeed, for finite, but small $\beta$, $\beta \ll 1$, the integral is approximately $(x/\beta)^{H-\delta}/(H-\delta)$. Thus, the derivative is negative over a finite interval $0<x<x_0$, where
\begin{equation}
x_0^{H-1} \approx \beta^{H-\delta} \frac{(\delta - H)(H-1)!}{m\delta }\frac{1}{1-\rho_0}
\end{equation}
According to the same argument as above, there is a  $\rho_0=\rho_c$  such that $g(x_0)=0$.

Let us now  consider the maximum degree heuristics, for which we have 
\begin{equation}
g(x) = 1 -\frac{x}{m}-\frac{\delta}{1-\beta^{\delta} } x^{\delta} \int_{x }^{x/\beta_0 }dte^{-t}t^{-\delta-1} \sum_{j=0}^{H-1} \frac{t^{j}}{j!}
\label{eq:dgdxMD}
\end{equation}
Here $\beta_0=m/m_0$, and the cutoff connectivity $m_0$ is given by Equation~\ref{eq:cutoff}, which, for $\beta\ll1$, reads \begin{equation}
 \beta_0=\rho_0^{1/(1+\delta)}.
 \end{equation}
 It is easy to check that 
\begin{equation} 
g(x=0)=\beta_0^{\delta}\equiv \rho_0^{\delta/(1+\delta)}.
\label{eq:gx0}
\end{equation}
 Furthermore, the derivate of $g(x)$ is 
\begin{equation}
g^{\prime}(x) \approx -\frac{1}{m}+ \frac{\delta}{(H-1)!} \frac{x^{H-1}}{H-\delta} \beta_0^{-H+\delta}
\end{equation}
Consequently, $g(x)$ is negative for $0<x<x_0$ where 
\begin{equation}
x_0^{H-1} \approx \beta_0^{H-\delta} \frac{(\delta - H)(H-1)!}{m\delta }\propto \rho_0^{(H-\delta)/(1+\delta)}
\end{equation}
Thus,  when decreasing $\rho_0$, the interval where $g(x)$ decreases shrinks as $\rho_0^{\alpha_1}$, with $\alpha_1 = \frac{H-\delta}{(H-1)(1+\delta)}$. At the same time, $g(x=0)=\rho_0^{\alpha_2}$, with $\alpha_2=\delta/(1+\delta)$.  For $\delta>1$ one has $\alpha_1>\alpha_2$, which suggests that $g(x)$ will cross the zero at some critical value $\rho_0=\rho_c$. And in contrary, for $\delta<1$ one has $\alpha_1<\alpha_2$, which means that $g(x)$  always remains positive as $\rho_0\rightarrow 0$, thus suppressing  critical behavior. Finally, repeating the arguments above, one can show  that introducing a connectivity cutoff for $\delta \le 1$will recover the criticality .

\bibliographystyle{plain}

\end{document}